\documentclass[twocolumn,epsfig,pre]{revtex4-1}

\usepackage{graphics}
\usepackage{graphicx}
\usepackage{epstopdf}
\usepackage{amssymb}
\usepackage{amsmath}
\usepackage{hyperref}
\usepackage{latexsym}
\usepackage{color}

\begin{document}

\renewcommand{\thefootnote}{\fnsymbol{footnote}} 
\renewcommand{\theequation}{\arabic{section}.\arabic{equation}}

\title{Ring and linear copolymer blends under confinement} 

\author{Lenin S. Shagolsem}
\email{slenin2001@gmail.com}
\affiliation{Department of Physics, National Institute of Technology Manipur, Imphal - 795004, India }
\date{\today}

\begin{abstract}

\noindent The behavior of dense mixtures of two topologically different diblock-copolymer (CP) chains, viz., linear(L)-CP and ring(R)-CP of same molecular weight which forms lamellae is studied under confinement by two non-selective substrates. The effect of varying interaction strength between L-CP and R-CP, from purely repulsive ({\it demixed state}) to weakly attractive ({\it mixed state}), on the morphology, domain size, chain conformations and distribution of chains in the film are investigated. In the demixed state, collective structure factor $S(q)$ shows a split of the predominant peak indicating the presence of two dominant length scales. While there is only one predominant peak in the mixed state, and hence a lamellar structure with single domain size. We show that the peak position $q*$ of $S(q)$ can be varied with the L/R interaction strength and thus allow one to control domain size by tuning L/R interaction strength without altering the chain size. We further characterize the chain size and illustrate that this domain size variation is a consequence of the variation in the size of L-CPs. Furthermore, results on the average instantaneous shape of R/L-CP reveal that their shapes are very different both in bulk and near the substrate, and R-CP assumes an oblate shape near the substrate. This shape/size difference leads to the segregation of R-CPs near the polymer-substrate interface and hence a relatively higher density of R-CPs at the interface. 

\end{abstract}

\maketitle


\section{Introduction}
\label{sec: intro}

In polymer materials the topology of polymer chains, besides nano-fillers, plays an important role in modifying physical properties such as strength, toughness, glass transition, and mechanical response, also it leads to interesting applications. For example, dendrimers and dendritic polymers (a subset of hyperbranched polymers) finds application ranging from coating to drug and gene delivery due to its unique structural properties- globular architecture and response to different solvent conditions.\cite{Gillies2005,Gao2004} 
Of much recent interest is the ring(R) or cyclic polymer, a topologically constrained polymer chain made by closing the chain ends of a linear polymer,\cite{mcleish2002,semlyen1994,polymeropoulos2017} which is not only relevant in physics (e.g., in understanding collapse of polymer gel), but also in biology (e.g., as a model system for chromatin folding and existence of chromosome territories).\cite{Halverson2014,hofmann2015} \\
 
Theory and experimental studies on ring polymer melts show that topological constraints influence both statics and dynamics. For example, sufficiently long ring polymers are shown to behave as compact object (i.e., size $\sim N^{1/3}$ with N chain length), also unexpected power-law stress relaxation is observed in entangled ring polymers, \cite{Vettorel2009,Kapnistos2008,Halverson2012,richter2015} and smaller diffusion coefficient~\cite{ozisik2002}. 
Interesting observations are made in case of ring-linear blends, e.g., it shows enhancement of miscibility due to the topological entropy gain of rings upon mixing with linear chains,\cite{khokhlov1996,sakaue2016} swelling of rings and unusual dynamics of rings in the matrix of linear chains,\cite{subramanian2008_PRE,shanbhag2017,jeong2017} and dramatic increase of viscosity of linear melts by adding rings~\cite{halverson2012}.  \
Recent efforts in understanding the equilibrium and out-of-equilibrium properties of ring polymers focus mainly on homopolymers. However, for technical applications one prefer block-copolymers. Block-copolymers are formed by chemically linking two or more immiscible polymer blocks which can microphase separate and form various self-assembled nanostructures.\cite{hamley_book,strobl_book,bates1999,ruzette2005} In this study, we refer to diblock-copolymers (CPs) which has only two types of immiscible blocks, say, A and B. \\

Copolymers find applications in thin-film technologies (e.g., nano-lithography, high-density information storage media, and energy storage),\cite{Hamley2003,Ozaydin2011,lodge2003,bates2017} also it is a promising candidate for making high efficiency organic photovoltaics since it form sharp interfaces and has the ability to control spatial organization of fullerene (charge acceptor) and thus enhance charge separation process.\cite{chai2007,bockstaller2005,darling2009} In a recent experimental study, R-CPs are used to control domain spacing which in the case of linear(L)-CPs relies on molecular weight and immiscibility as a parameter to control, and thus establish the fact that chain architecture is also an important tool for tuning domain spacing and other features.\cite{Poelma2012} 
Ring CPs can be prepared for example by a combination of atom transfer radical polymerization and click cyclization.\cite{Eugene2008} \\

Although both L-CP and R-CP show the same nature of order-disorder transition they differ in many respect due to the difference in their topology: two chain ends for L-CP Vs no chain ends for R-CP. For R-CP there are two connections between the blocks in the same chain and, therefore, it is expected that its properties should differ from those of L-CP. Earlier studies regarding microphase separation, conformational properties, and phase behavior,\cite{Qian2005_Macromol,Lecommandoux2004_Macromol,Jo1999_JCP,Marko1993_Macromol,Vlahos1995_Macromol,Pakula1988_Macromol} and rheology~\cite{Mozorov2001_Macromol} of pure R-CP melts focuses on the systems in bulk. And to our best knowledge the behavior of L-CP and R-CP blends in thin-film geometry has received little attention. R-CPs in contrast to its linear counterpart show differences due to its topological constraints, e.g., relatively higher $\chi_{_{\rm ODT}}$ due to smaller concentration fluctuations, and relatively small chain size.\cite{Qian2005_Macromol,Jo1999_JCP,Marko1993_Macromol} Blends of R/L-CPs represents an interesting system to explore the effect of mixing two topologically different chains, where each type can form self-organized structures, and its consequences on the global structure formation.\\

In this paper, by considering generic bead-spring polymer model, we investigate the behavior of thin-films made of ring-linear CP blends under confinement by means of molecular dynamics simulations. We consider unknotted and non-concatenated rings, and the confining walls are non-selective thus forming a vertically oriented lamellae. In this study, we focus our attention on the studies of morphologies of the R/L-CP blends under different ring-linear interaction parameters where we characterize the domains sizes and show that it is possible to tune the domain size by controlling the interaction strength. Further, we discuss the chain conformation properties and characterize its instantaneous shapes, and finally segregation/distribution of ring/linear polymers in the film is discussed. \\

Our paper is organized as follows. Model and simulation details are described in section~\ref{sec: model-description}. In section~\ref{sec: results}, the results are presented, and finally in section~\ref{sec: summary} we summarize and discuss our results. 



\section{Model and simulation details}
\label{sec: model-description}

Blends of linear (L) and ring (R) copolymers (50:50 mixture) is modeled via coarse-grained polymer chains. Here, the polymer chains are represented by coarse-grained monomers or beads connected with springs (Kremer-Grest model).\cite{kremer_JCP_92} All the pairwise interactions in the system is modeled via Lennard-Jones (LJ) potential,
\begin{equation}
U_{\tiny{_{\rm LJ}}}(r) = 4\epsilon\left[(\sigma/r)^{12}-(\sigma/r)^{6}\right]~,
\label{eqn: LJ-potential}
\end{equation}
with $r$ the separation between a pair of particles, $\epsilon$ depth of the potential, and $\sigma$ monomer diameter. The LJ potential is cut-off at a distance $r_c$ and it is shifted to zero to avoid discontinuity of the force at the cut-off.
The monomer connectivity along the chain is assured via finitely extensible, nonlinear, elastic (FENE) springs~\cite{grest_kramer_PRA_33} represented by the potential, 
\begin{equation}
U_{\tiny{_\text{FENE}}} = \left\{
 \begin{array}{l l}
 -\frac{kr_0^2}{2}\ln\left[1-\left(r/r_0\right)^{2}\right]~, & \quad r<r_0 \\ 
 \infty~, & \quad r\ge r_0
 \end{array} \right.
\label{eqn: fene-potential}
\end{equation}
where $r$ is the separation of neighboring monomers in a chain. The spring constant $k=30\epsilon/\sigma^2$ and the maximum extension between two consecutive monomers 
along the chain $r_0=1.5\sigma$. The above values of $k$ and $r_0$ ensure that the chains avoid bond crossing and very high frequency modes.\cite{kremer_JCP_92}  
All the physical quantities are expressed in terms of LJ reduced units where $\sigma$ and $\epsilon$ are the basic length and energy scales respectively. The reduced temperature $T$ and time $t$ are defined as $T = k_{_B}T_0/\epsilon$ and $t = t_0/\tau_{_{\rm LJ}}$, where $\tau_{_{\rm LJ}}=\sigma\sqrt{m/\epsilon}$ represents the LJ time unit, and $k_{_B}$, $T_0$, $t_0$, and $m$ are the Boltzmann constant, absolute temperature, real time, and mass, respectively. \\

For simplicity, we imagine that both L-CP and R-CP chains comprise of two types of beads, say, A and B. The A-B interaction is modeled by a purely repulsive LJ potential cut-off at potential minimum ($r_c=2^{1/6}\sigma$), while we allow attraction between the same type of monomers ($r_c=2\sigma$). And we vary the interaction between L-CP and R-CP chains, i.e. between monomers of type-A (or type-B) of L-CP and R-CP, from purely repulsive to weakly attractive, while A-B interaction is repulsive irrespective of the chain types. Various cut-off distances of the LJ potential for different pairs in the system is summarized in table~\ref{table: cut-off-radii}. \

\begin{table}[h]
\caption{Pair-wise interaction range among the species}
\centering
\begin{tabular}{|c|c|c|}
\hline
Interaction between & Cut-off radius $r_c$/$\sigma$ & Nature of interaction \\
\hline
A - B & $1.12$ & repulsive \\
A - A or B - B & $2.0$ & attractive \\
A/B - wall & $1.12$ & repulsive \\
Linear - Cyclic & $1.12-2.0$ & repulsive~--~\\
& & attractive \\
\hline
\end{tabular}
\label{table: cut-off-radii}
\end{table}

The molecular dynamics (MD) simulations were carried out using Langevin dynamics\cite{allen-book,frenkel-smit-book} where the equations of motion are given by  
\begin{equation}
 m_i \frac{d^2 {\bf r}_i}{dt^2} + \gamma \frac{d {\bf r}_i}{dt} = -\frac{\partial U}{\partial {\bf r}_i} + {\bf f}_i(t)~,
 \label{eqn: langevin}
\end{equation}
with ${\bf r}_i$ and $m_i$ the position and mass of particle $i$, respectively, $\gamma$ the friction coefficient which is taken to be the same for all particles, $U=U_{\tiny{_{\rm LJ}}}+U_{\tiny{_\text{FENE}}}$ is the potentials acting on monomer $i$ of the polymer chain, and ${\bf f}_i$ are random external forces which follows the relations: $\left\langle {{\bf f}_i}(t)\right\rangle=0$ and $\left\langle {{\bf f}_i}^\alpha(t){{\bf f}_j}^\beta(t')\right\rangle=2\gamma m_i k_{_B}T\delta_{ij}\delta_{\alpha\beta}\delta(t-t')$ where $\alpha$ and $\beta$ denotes the Cartesian components. 
The friction coefficient can be written as $\gamma=1/\tau_d$, with $\tau_d$ the characteristic viscous damping time which we fixed at $50$ and it determines the transition from inertial to overdamped motion (due to collisions with molecules of the implicit "solvent") in the dilute system limit. At the rather high density in the present simulation, the damping due to collisions between the particles dominates and the transition to the overdamped regime takes place on much faster time scales. The chosen value of $\gamma$ gives correct thermalization at the selected temperature $T=1$ for our study. The equations of motion are integrated using velocity-Verlet scheme with a time step of $\delta t_0 = 0.005 \tau_{_{\rm LJ}}$. \\ 

By considering symmetric CP chains consisting of $N=48$ monomers each chain, we prepare thin-films of L/R-CP blends by mixing L-CPs and R-CPs in equal amount (with 1500 chains in total) in a simulation box of dimensions: $L_x=L_y=50\sigma$, and $L_z=25\sigma$ in X-, Y-, and Z- directions, respectively. The simulation box is periodic along both X- and Y- axes, while it is non-periodic in Z- direction due to the presence of explicit atom walls. The surface of the confining walls are located at $z=0$ and $z=25$, and each wall consists of 2500 atoms. The rigid wall atoms (with diameter $\sigma$) are arranged in square lattice with nearest neighbors separated by $\sigma$. For the given box dimensions, the reduced number density of the pure system is $\rho^\ast\approx0.96$ (polymer melt regime), and the volume fraction $\phi\approx 0.6$. 
The simulations are carried out under fixed NVT condition using the open source MD simulation package LAMMPS.\cite{lammps} We set $\epsilon=0.5$ for all the pair interactions and both monomer diameter and mass are set to unity; however, different type of pairs have different interaction cut-off radii listed in table~\ref{table: cut-off-radii}. 
We prepare our systems for investigation through the following protocol: First, disordered systems at high temperature are prepared and then quench the systems at T=1.0, which is well below the order-disorder transition (ODT) temperature, followed by relaxation for $5\times 10^6$ MD steps, where well ordered lamellar structure is formed. And further productions runs are carried out for another $1\times 10^7$ MD steps, where various measurements are performed. Since the confining walls are non-selective a vertically oriented lamellar structure is formed.\cite{sommer_hoffmann_jcp1999} 

\begin{figure}[ht]
\begin{center}
\includegraphics[width=0.5\textwidth]{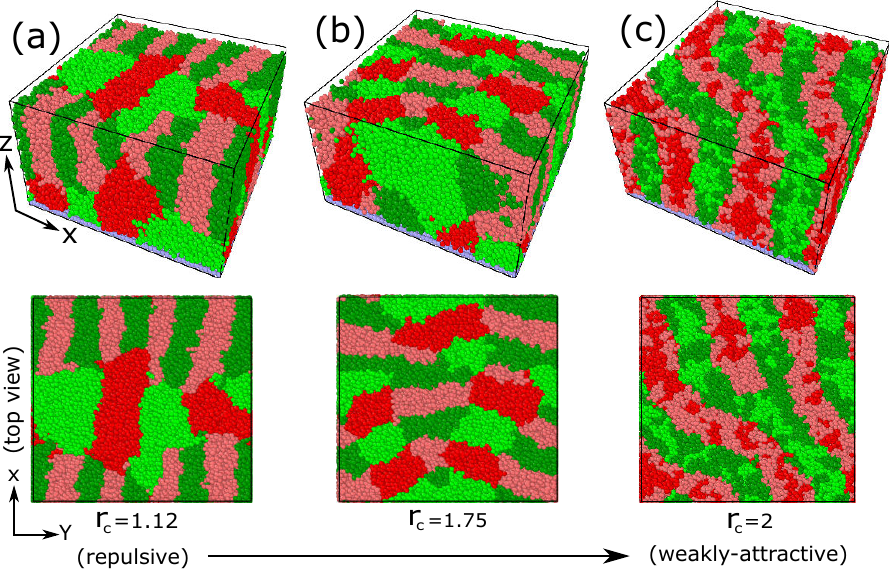}
\caption{Simulation snapshots of the L-CP and R-CP blends for non-selective walls at different values of L-R interaction parameters ranging from repulsive to weakly-attractive as indicated in the figure. Color scheme: Bright (or dull) red and green beads represents L-CP (or R-CP). For clarity the upper wall is removed, while the bottom wall is kept.}
\label{fig: simulation-snapshot} 
\end{center}
\end{figure}

\section{Results and Discussion}
\label{sec: results}

In figure~\ref{fig: simulation-snapshot}, we display the equilibrium morphologies of the blend at different values of R/L interaction strength characterized by the value of $r_c$ where we vary it in the range: $1.12~({\rm repulsive}) \le r_c \le 2~({\rm weakly-attractive})$. As one can see, for $r_c=1.12$ the R/L-CPs are segregated at large length scales and we refer to this as {\it demixed state}, whereas for $r_c=2$ they are miscible and we refer to this as {\it mixed state}. For intermediate values of $r_c$, we see partially mixed states, e.g., see figure~\ref{fig: simulation-snapshot}(b) for $r_c=1.75$. Furthermore, from the visual inspection of figure~\ref{fig: simulation-snapshot}, it is clear that in the demixed state there are two length scales corresponding to different domain sizes of the lamellae formed by L-CP and R-CP, while in the mixed state there is only one domain size. In order to characterize the average domain size, we first calculate the collective structure factor which is then compare with the results from the straightforward method of box counting.\cite{binder1988,sommer_hoffmann_jcp1999}  


\subsection{Domain size} 

Below the ODT the lamellar structure appears and for which the average domain spacing is given by $\lambda=2\pi/q^*$, where $q^*$ is the wave vector for which the maximum value of the structure factor $S(q)$ is obtained, and thus the periodicity of the composition fluctuation in the system is $\lambda$. One can directly calculate $S(q)$ from the simulations data of CP system as follows. To each monomer we assign a spin-type variable $s({\bf r_i})$, where ${\bf r_i}$ is the position vector of $i^{\rm th}$ monomer. Here, $s({\bf r_i})=+1$ for A-monomer and $s({\bf r_i})=-1$ for B-monomer. Now, the structure factor is obtained as
\begin{equation}
S(q)=\frac{1}{V}\sum_{i,j} {\left\langle s({\bf r_i})s({\bf r_j}) \right\rangle} e^{i{\bf q}({\bf r_i - r_j})}~, 
\label{eq: sq} 
\end{equation}  
where $V$ is volume of the system. It is important to point out that since our system is a thin-film (i.e, finite system) only a discrete set of q-vectors are physically meaningful and it depends on the dimensions of the system. 
\begin{equation}
{\bf q} = \frac{2\pi}{L_x}(n_x,n_y,n_z)~,
\end{equation}
with $0 \le n_i \le L_i$ for $i=x,y,z$. Using these allowed q-vectors we calculate $S(q)$ for the blends, and to have a good statistics we average over about 200 -- 300 configurations. For reference, we also calculate $S(q)$ for thin-films consisting of only L-CPs and only R-CPs. \\

\begin{figure}[ht]
\begin{center}
\includegraphics[width=0.45\textwidth]{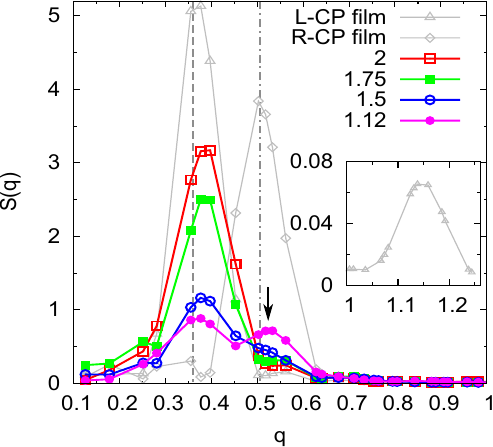} 
\caption{The collective structure factor $S(q)$ of the reference clean systems and that of the R/L-CP blends at different values of L-R interaction strength (i.e., different values of $r_c$) indicated in the figure. The positions of the maximum of $S(q)$ for pure L-CP and R-CP films are indicated by the vertical dashed and dashed-dotted lines at $q^*_{_L}\approx 0.36$ and $q^*_{_R}\approx 0.5$, respectively. Arrow vertically pointed downward at $q\approx 0.525$ indicates the peak position. Inset figure shows the higher-order Bragg's peak for pure L-CP film. 
}  
\label{fig: sq} 
\end{center}
\end{figure}

In figure~\ref{fig: sq}, we show $S(q)$ for the R/L-CP blends and that for the clean reference systems. For the reference L-CP system the maximum of $S(q)$ is located at $q^*_{_L}\approx 0.36$ which corresponds to a domain spacing of $\lambda_{_L}\approx 17.5$, and that for the reference R-CP system it is located at $q^*_{_R}\approx 0.5$ and correspondingly $\lambda_{_R}\approx 12.5$. It is interesting to note that $S(q)$ for R/L blend in the demixed state ($r_c=1.12$ in the figure) have two peaks of roughly same intensity located very close to $q^*_{_L}$ and $q^*_{_R}$. As seen above, in the demixed state, L-CP and R-CP are segregated on large length scales and thus the observed two peaks corresponds to the contributions from each clean phase with distinct domain sizes. However, when we introduce attractive interaction ($1.12<r_c<2$) peak around $q^*_{_L}$ survives indicating mixing. The observed trend is that as we increase the interaction strength the position of the peak moves towards the larger value of q-vector with the maximum value achieved in the mixed state ($r_c=2$) where $q^*_{_{LR}}\approx 0.39$ and thus $\lambda_{_{LR}}\approx 16$. From this observation it is clear that in R/L-CP blends it is possible to control the domain size of the lamellae by tuning the interaction parameters alone. 

\begin{figure}[ht]
\begin{center}
\includegraphics[width=0.5\textwidth]{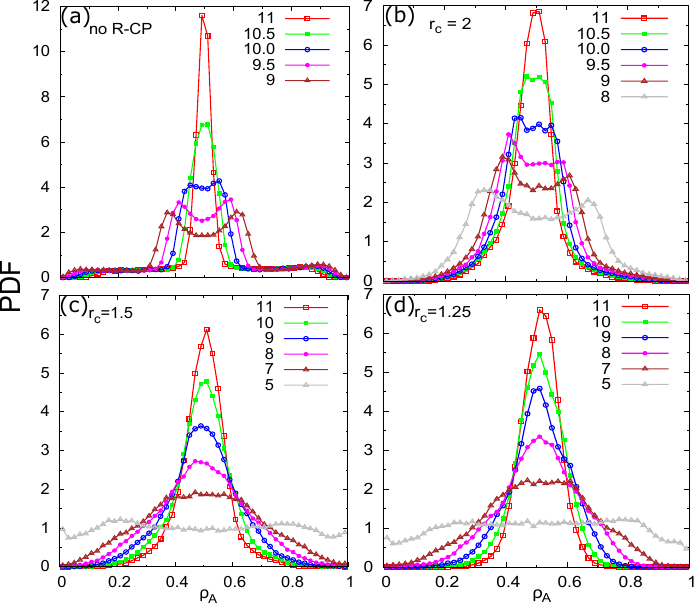} 
\caption{PDF of concentration of A-monomers $\rho_{_A}$ obtained from the box counting method: fig.~(a) for reference L-CP thin-films, and figs.~(b)-(d) for L/R-CP blends at different values of interaction strength indicated by the value of $r_c$.} 
\label{fig: pdf-concA-box-counting} 
\end{center}
\end{figure}

We now proceed to compute the domain size by a more direct approach, i.e., box counting method.\cite{binder1988} In an earlier work by Hoffmann {\it et al.},\cite{sommer_hoffmann_jcp1999} box counting method was successfully applied to estimate the average domain size of the lamellar structure. 
In this method, the concentration of one component of the copolymer, say, A-monomers $\rho_{_A}$ in a sphere of radius $r$ drawn at a random position within the simulation box is calculated. For spheres at different locations different values of $\rho_{_A}$ will be obtained and hence a distribution function $P(\rho_{_A})$. For sphere radius much larger than the average domain size $P(\rho_{_A})$ is a Gaussian distribution function. 
However, for sphere dimension of the order of domain size the spatial extension of the concentration fluctuation exceeds the volume and thus regions containing only A-monomers (or B-monomers) dominates leading to a bimodal distribution function. The value of average domain size is then given by the sphere radius $r^*$ at which $P(\rho_{_A})$ changes its shape from unimodal to bimodal. 
In our calculations, to obtain the concentration of A-monomers we calculate the number fraction, i.e., $\rho_{_A}=n_{_A}/(n_{_A}+n_{_B})$ with $n_{_{A/B}}$ the total number of A/B-monomers enclosed by the sphere, instead of volume fraction, and obtain the probability density function (PDF). Since $\rho_{_A}$ is a number fraction it can vary in the range $0\le \rho_{_A}\le 1$ and hence the peak positions can be different from the one calculated using the volume fraction. 
In figure~\ref{fig: pdf-concA-box-counting} we show the PDF of the concentrations of A-monomers for both reference L-CP system and R/L-CP blends calculated at different values of $r$. 
In order to calculate the distribution, we obtain $\rho_{_A}$ from 500 spheres drawn at random positions and for better statistics sampled around 300 different configurations. 
As shown in figure~\ref{fig: pdf-concA-box-counting}(a), for the reference L-CP thin-films the distribution has a single peak for $r>9.5$ and at $r=r^*_{_L}\approx 9.5$ the single peak splits into two distinct peaks and thus is roughly the average thickness of the lamellae, also the value is close to the one obtained from $S(q)$. For R/L-CP blend in the mixed state, see figure~\ref{fig: pdf-concA-box-counting}(b), two distinct peaks appears at $r^*_{_{LR}}\approx 8$ which exactly matches with the result from $S(q)$, i.e., $r^*_{_{LR}}=\lambda_{_{LR}}/2$. 
And in the demixed/partially mixed states we get $r^*_{_{LR}}\approx 5$ (slightly smaller than $\lambda_{_R}/2$) and the value is smaller than $r^*_{_{LR}}$(mixed state) and thus corresponds to the thickness of the lamellae formed by R-CPs. 
Although there are regions of L-CPs or R-CPs in the demixed/partially mixed states we see two peaks for $r^*_{_{LR}}=5$ only because for $r>r^*_{_{LR}}$ the spatial extension of concentration fluctuations of R-CPs is smaller than the sphere volume and smears out the contribution due to L-CPs and thus a unimodal distribution is observed. So far we have discussed the overall structure (i.e., domain size) of the lamellae form by the R/L blends and in the following we discuss the chain conformational properties.   
 
 
\subsection{Chain conformations}
 
\begin{figure}[ht]
\begin{center}
\includegraphics[width=0.45\textwidth]{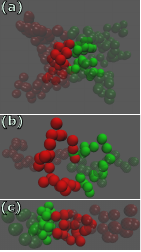} 
\caption{Typical conformation of a R-CP in the bulk of R/L-CP blend for $r_c=2$ (mixed state). Here, a single R-CP in the bulk is selected and display together with neighboring L-CPs whose center-of-mass falls within the $R_g$ of the R-CP, and rest of the chains are not shown. For clarity R-CP is shown in solid red and green beads and L-CPs are made semi-transparent. Notice that the spatial extension of ring polymer is much lesser compare to the linear chains. In figs.(b) and (c), notice the threading of L-CP through the R-CP.} 
\label{fig: snap1} 
\end{center}
\end{figure}

In order to understand the conformational properties of CP chains in L/R-CP blends, we calculate the mean-square radius of gyration 
\begin{equation} 
\left\langle R^2_g \right\rangle = \left\langle \frac{1}{N}\sum_{i=1}^N({\bf r_i-r_{_{cm}}})^2\right\rangle~,
\end{equation} 
with $N$ the number of monomers per chain, ${\bf r_i}$ and ${\bf r_{_{cm}}}$ the monomer position and center-of-mass of the chain, respectively, and angular bracket represents the ensemble average. Since we are considering a confined system we calculate $\left\langle R^2_g\right\rangle$ as a function of height/distance $h$ from the walls in order to see the influence of wall on the chain conformations. To obtain the layer resolved $\left\langle R^2_g\right\rangle$ we first subdivide the film into thin slices along the Z-axis and count the chains whose center-of-mass belongs to a particular slice and then the $\left\langle R^2_g\right\rangle$ is calculated and averaged for each slice. We averaged over 300 configurations for good statistics. A typical R-CP conformation in the mixed state of the R/L-CP blend is shown in figure~\ref{fig: snap1} where it is seen that the R-CPs adopt a much more compact structure compare to the linear counterpart. It is clear from figure~\ref{fig: snap1}(b) that the L-CP passes through R-CP and it is interesting to see such threading of linear chains through rings, also see figure~\ref{fig: snap1}(c). \\

\begin{figure}[ht]
\begin{center}
\includegraphics[width=0.5\textwidth]{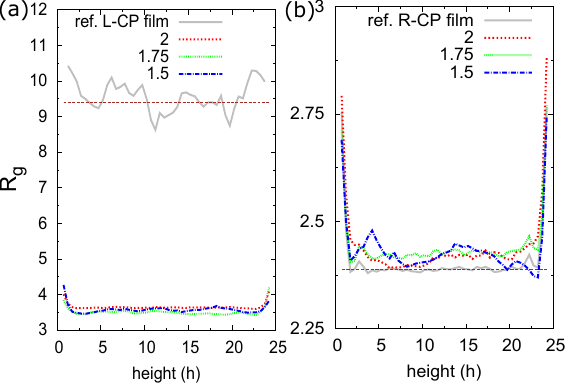} 
\caption{$R_g=\sqrt{\left\langle R^2_g\right\rangle}$ as a function of height from the walls for L/R-CP blends at three different values of $r_c$ indicated in the figure shown together with the reference systems. Fig.~(a) for reference L-CP film and L-CPs in the blend, fig.~(b) for reference R-CP film and R-CPs in the blend. Walls are located at $h=0$ and $h=25$. Horizontal dashed lines at $y=9.4$ (fig.~a) and $y=2.38$ (fig.~b) corresponds to the value of $R_g$ in bulk for the reference systems.} 
\label{fig: rg} 
\end{center}
\end{figure}

\begin{figure}[ht]
\begin{center}
\includegraphics[width=0.45\textwidth]{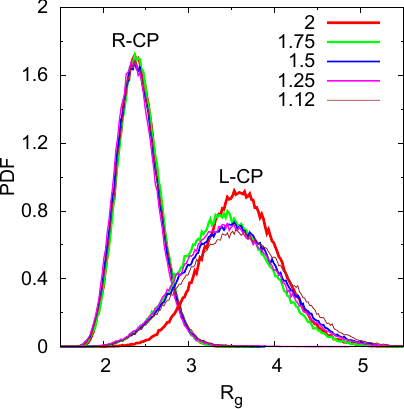} 
\caption{PDF of $R_g$ for the ring and linear CPs in the R/L-CP blend system shown for different values of R/L interaction strength indicated by the values of $r_c$ in the figure.} 
\label{fig: rg-pdf} 
\end{center}
\end{figure}

In figure~\ref{fig: rg}, we display the layer resolved $R_g=\sqrt{\left\langle R^2_g\right\rangle}$ for the R/L-CP blends at different values of R/L interaction strength. For the reference systems, we find $R_g\approx 9.41$ (ref.~L-CP film) and $2.38$ (ref.~R-CP film) in bulk, see figures~\ref{fig: rg}(a) and (b) respectively. In the film, we define bulk region as the region at least a distance of $d=5\sigma$ away from the confining surfaces, where for distance larger than $d$ the influence of confinement on the Z-component of $R_g$ (i.e., normal to the confining surface) vanishes (not shown here). The value of $d$ for R-CPs is roughly half of the L-CP.  It is interesting to note that the size of L-CPs in the blends decreases by a factor of 2.6 approximately, while for R-CPs the $R_g$ is slightly increased compare to the reference R-CP. So, in the R/L blends the linear chains shrink significantly, whereas ring chains swell slightly. This leads to an overall decrease in the domain size of the lamellae formed as reflected in the $S(q)$. In table~\ref{table: rg-values}, we show the value of $R_g$ for the blends at different values of interaction parameter. Although the change of size within the considered interaction range is small its overall effect on the domain size is significant as indicated by the significant shift in the peak position of $S(q)$. Note that the swelling of ring polymers upon addition of small molecules/solvents and linear polymers in the case of homopolymers has been observed in earlier studies,\cite{subramanian2008_PRE,shanbhag2017,jeong2017} however it is reported that the nature of swelling is not clear. And in a recent study by Jeong and Douglas it is shown that the swelling arises from altered self-excluded volume interactions which amplifies in the entangled regime.\cite{jeong2017} This suggest that, for our R/L-CP blends, swelling of R-CPs may be amplified for longer chain lengths. However, detail study on the internal structure and the nature of swelling of R-CPs at various degrees of polymerization is left to future investigation. 
The PDF of $R_g$ for the ring and linear CPs of the blend at different ring-linear interaction strength are shown in figure~\ref{fig: rg-pdf}. Notice that for R-CP the distribution is independent of the interaction strength considered, while for L-CP there is a significant shift in the peak position. This points to the fact that the observed variation of domain size is mainly due to change in the size of L-CPs. \ 

\begin{table}[h]
\caption{Measured values of $R_g$ in the bulk for L-CP and R-CP in R/L-CP blends for different values of linear-ring interaction strength characterized by the cut-off distance $r_c$.}
\centering
\begin{tabular}{|c|c|c|}
\hline
$r_c$/$\sigma$ (cut-off between & L-CP & R-CP \\ 
L-CP and R-CP) &~~&~~\\
\hline
~~~2.0~~~ & $~~3.6399	\pm 0.444~~$ & $~~2.4095 \pm 0.225~~$ \\
1.75 & $3.4892 \pm 0.510$ & $2.4244	\pm 0.223$ \\
1.5 & $3.5724	\pm 0.548$ & $2.4172	\pm 0.241$ \\
1.25 & $3.5703 \pm 0.558$ & $2.4040	\pm 0.266$ \\
1.12 & $3.6291 \pm 0.585$ & $2.4462 \pm 0.265$ \\
\hline
\end{tabular}
\label{table: rg-values}
\end{table}

\begin{figure}[ht]
\begin{center}
\includegraphics[width=0.5\textwidth]{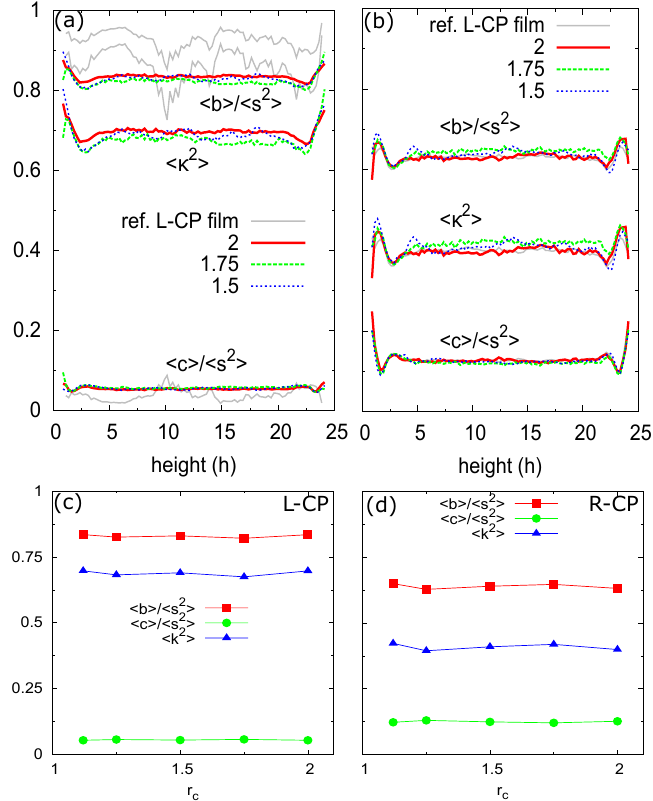} 
\caption{Normalized shape parameters ${\left\langle b\right\rangle}/{\left\langle s^2 \right\rangle}$ (asphericity), ${\left\langle c\right\rangle}/{\left\langle s^2 \right\rangle}$ (acylindricity), and ${\left\langle \kappa \right\rangle}$ (relative shape anisotropy) of the linear and ring CPs for the clean reference systems and that in R/L-CP blends at three different values of R/L interaction strength. Figs.~(a)-(b): Shape parameters as a function of height from the confining walls, and figs.~(c)-(d): mean values in the bulk.} 
\label{fig: shape-param} 
\end{center}
\end{figure}

Since linear and ring polymers are topologically different it is important to investigate the differences in their instantaneous shapes and how the confinement alter their instantaneous shapes. It is well established that the average instantaneous shape of a linear polymer chain is not spherical and it is characterized by means of the radius of gyration tensor {\bf S} as a shape measure.\cite{ch3_doros1985,ch3_solc_stockmayer1971,ch3_solc1971} Here, a principal axis system is chosen where {\bf S} is in diagonal form and the eigenvalues $\lambda_1^2, \lambda_2^2$ and $\lambda_3^2$ are such that $\lambda_1^2 \le \lambda_2^2 \le \lambda_3^2$. This construction leads us to the picture of an ellipsoidal shape with eigenvalues as the semi-axes of the ellipsoid. The shape anisotropy of a chain is characterized in terms of the following shape parameters 
\begin{align}
 b &= \lambda_3^2 - \frac{1}{2} (\lambda_1^2+\lambda_2^2)~,\nonumber \\
 c &= \lambda_2^2 - \lambda_1^2~, ~\text{and} \nonumber \\
 \kappa^2 &= (b^2+\frac{3}{4}c^2)/s^4~,
 \label{eqn:shape-param}
\end{align}

\noindent where $b$, $c$, and $\kappa^2$ are asphericity, acylindricity, and relative shape anisotropy, respectively, and $s^2=\lambda_1^2+\lambda_2^2+\lambda_3^2 = \text{tr}({\bf S})$ is the squared radius of gyration of the chain. The quantity $b$ and $c$ measure the deviations from spherical and cylindrical shapes, respectively, where $b=0$ and $c=0$ for perfect sphere and cylinder, respectively. \\

\begin{figure}[ht]
\begin{center}
\includegraphics[width=0.4\textwidth]{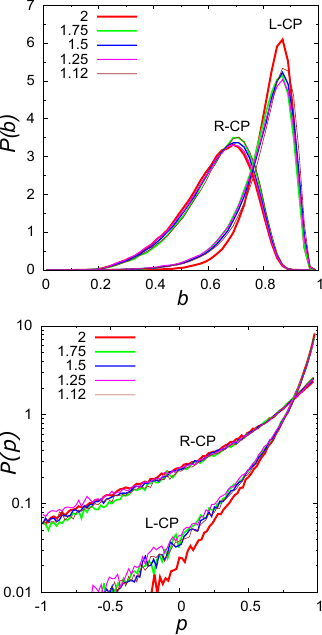} 
\caption{Normalized distribution of asphericity parameter (upper panel) and prolateness parameter (lower panel) for R-CP and L-CP of the R/L-CP blend at different ring-linear interaction strength.} 
\label{fig: shape-param-dis} 
\end{center}
\end{figure}

\begin{figure}[ht]
\begin{center}
\includegraphics[width=0.47\textwidth]{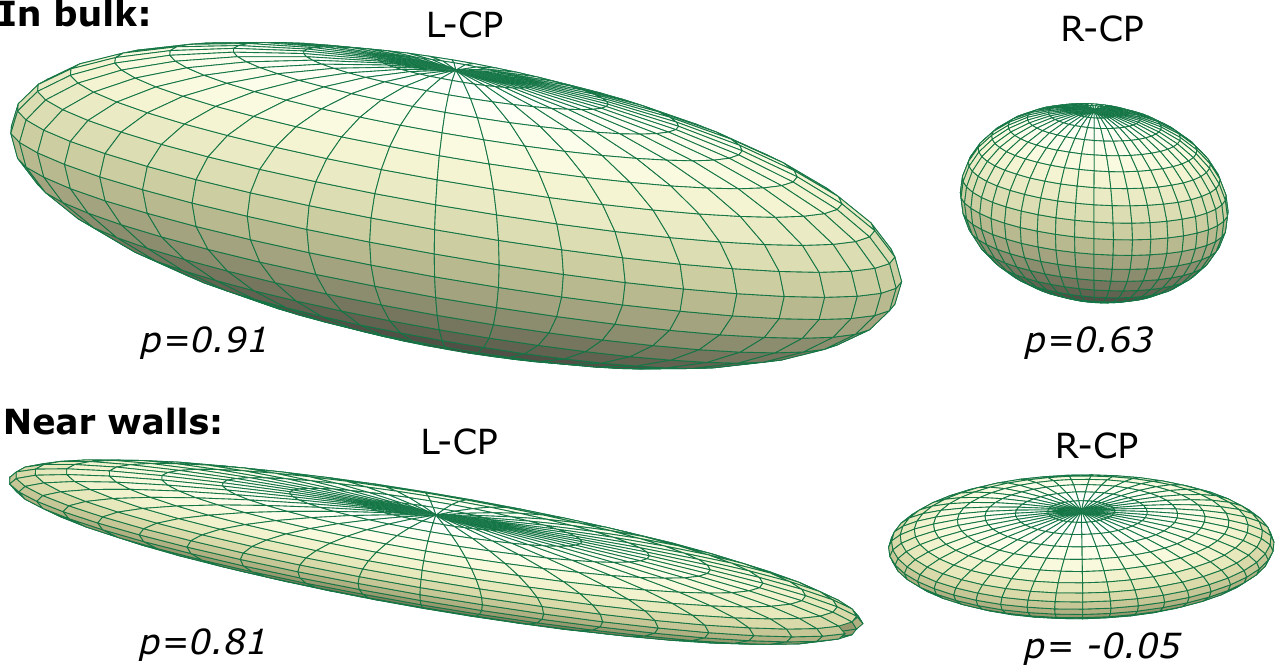} 
\caption{Instantaneous shapes of L-CP and R-CP in the bulk and near the walls drawn using the average eigenvalues of the gyration tensor in the mixed state (i.e.~$r_c=2$) are shown along with prolateness parameter $p$. Both linear and ring CPs near the walls looks more flat and elongated, see text for detail, however as indicated by the parameter ($p<0$) only R-CP has oblate shape near the walls. Figures are drawn at the same scale in the respective regions only. See figs.~\ref{fig: shape-param}(c) and (d) for the shape parameter values.} 
\label{fig: ellipsoid} 
\end{center}
\end{figure}

In order to see the behavior of shape parameters in various regions in the film we plot the layer resolved values for both polymer types along with the reference clean systems, see figure~\ref{fig: shape-param}. As we can see in figure~\ref{fig: shape-param}(a), for L-CP in the blend, overall the values of $b$ and $\kappa^2$ are decreased, while $c$ is slightly increased compared to the L-CP in the reference system. This indicates that L-CPs in the blends, whose size shrinks in the blend, see figure~\ref{fig: rg}, are relatively less cylindrical and more spherical. Although the shape is still ellipsoidal, L-CPs in the blend are more compact compared to the chain in the reference system. Due to confinement an increase in the value of shape parameters is observed as we approaches the walls.  On the other hand, for R-CPs no large change in the value of shape parameters (relative to the values for reference system) is observed upon mixing with L-CPs, figure~\ref{fig: shape-param}(b). However, in comparison with L-CPs in the blend, the value of $b$ (or $c$) is significantly smaller (or larger) and hence a much smaller value of $\kappa^2$. Thus, on average R-CPs adopt a much more rounded and hence a compact conformation. The mean values of the shape parameters in bulk at different values of R/L interaction strength are shown in figures~\ref{fig: shape-param}~(c) and (d).\
 
To see the full spectrum of various chain shapes in the film, in figure~\ref{fig: shape-param-dis} we compare the distributions of asphericity parameter $P(b)$ and prolateness parameter $P(p)$  with  
\begin{equation}
p = \frac{(2\lambda_1-\lambda_2-\lambda_3)(2\lambda_2-\lambda_1-\lambda_3)(2\lambda_3-\lambda_1-\lambda_2)}{2(\lambda_1^2+\lambda_2^2+\lambda_3^2-\lambda_1\lambda_2-\lambda_2\lambda_3-\lambda_3\lambda_1)^{3/2}}~,
\label{eq: prolateness}
\end{equation} 
where the prolateness parameter $p=1$ for perfectly prolate shape and $p=-1$ for perfectly oblate shape. In case of R-CPs, peak of the distribution $P(b)$ systematically moves to smaller value of $b$ in going from demixed state ($r_c=1.12$) to mixed state ($r_c=2$), while for L-CPs the peak position roughly remains the same. On the other hand, distribution $P(p)$ for R-CPs extends the entire range $p:[-1,1]$ and we see that a significant fraction of R-CPs have $p<0$ indicating that both prolate and an oblate shapes are present. Whereas, for L-CPs, the fraction of chains with $p<0$ are negligible and hence almost all the L-CPs are prolate in shape. 
To check this we compare chain shapes in the bulk and near the walls by drawing ellipsoids using the average eigenvalues of the gyration tensor in the respective regions as shown in figure~\ref{fig: ellipsoid}. In consistent with similar studies on homopolymer ring/linear systems,\cite{subramanian2008,iyer2007,bohn2010} clearly we see that the average instantaneous shape of R-CPs are different from the linear counterpart in both bulk and near-wall. 
The prolateness parameter $p$ is +ve for L-CPs in both bulk/surface regions, whereas for R-CPs it is +ve/-ve in the bulk/surface region. Thus, R-CPs assumes prolate shape in the bulk, whereas it assumes an oblate shape near the confining surfaces. L-CPs remains strongly prolate both in bulk and surface regions. However, irrespective of the shape difference, overall R-CPs are more compact and thus have a higher local segmental density. In the following we discuss the distributions of chains in the film. 


\subsection{Distribution of Chains} 
\label{subsec: chain-distribution}

The distribution of center-of-mass of the chains in the film is investigated by calculating the laterally averaged concentration profile of linear and ring polymers. To obtain the concentration profile we first subdivide the film into thin slices and calculate the fraction of L-CP (or R-CP) in each layer.  \\ 

\begin{figure}[ht]
\begin{center}
\includegraphics[width=0.45\textwidth]{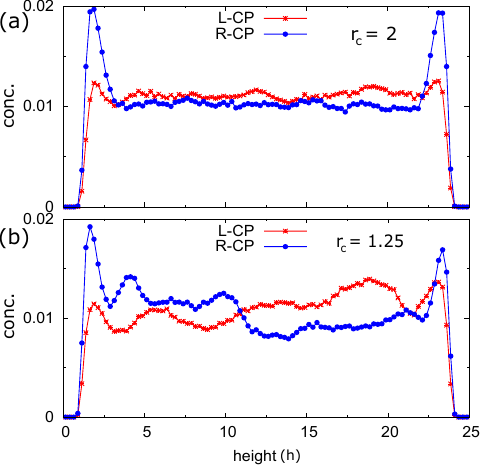} 
\caption{Distribution of L-CP and R-CP in the film for the R/L-CP blend in the mixed state, fig.~(a), and partially mixed state, fig.~(b). Notice the segregation of R-CPs at the polymer-wall interface.} 
\label{fig: concentration-profile} 
\end{center}
\end{figure}

In figure~\ref{fig: concentration-profile}, we compare the distribution of ring and linear CPs in the film for two different values of R/L interaction strength corresponding to mixed state and partially mixed state. It is interesting to note that in both cases (mixed and partially mixed states) the concentration of R-CPs at the polymer-wall interface is higher indicating that R-CPs are segregated at the interface. Such segregation at the polymer-wall interface is observed in other composite systems, e.g., polymer and nanoparticle mixtures where the polymer-induced depletion attraction drives the non-selective nanoparticles to the walls.\cite{shagolsem-mats2011,balazs-science2006,gupta-nmater2006,krishnan-condmat2007,balazs-PRL2003,balazs-macromol2003,mackay-PRL2007,mackay-JCP2008,pryamitsyn2006} Since the depletion forces arise as a result of mixing polymer with bigger particles like nanoparticles/colloids.\cite{asakura1954,asakura1958} We can imagine similar situation here also since from the above size and shape analysis and visual inspection we know that R-CPs adopts a much more compact structure, see figure~\ref{fig: ellipsoid}, and thus behave as a big soft-and-patchy particles, say, soft Janus particles, and thus have the tendency to go to the polymer-wall interface. 


\section{Summary}
\label{sec: summary}

In this article, we have presented computer simulations study on the equilibrium behavior of 50:50 mixture of two topologically different diblock-copolymers, viz.~ring and linear chains of same polymerization index, which form lamellar phase under confinement by non-selective substrates. In particular, we have investigated the effect of varying ring-linear interaction strength on the morphology, effect on chain conformations and instantaneous shapes, and distribution behavior of chains in the film. The results obtained are summarized and discuss below.\

Since the confining walls are non-selective the lamellae are oriented normal to the walls. We observed that for purely repulsive ring-linear interaction the ring and linear CPs macro-phase separate forming regions of ordered L-CPs and R-CPs which have different domain sizes where the domain size is larger for L-CP. While in the case of weakly-attractive (i.e., R-L interaction strength same as interaction between the like monomers) the R-CPs and L-CPs are completely miscible and lamellar structure with single domain size is formed. Partially mixed states are observed for the intermediate values of interaction strength. We observed shifting of the peak position of the collective structure factor $S(q)$ to a higher value of $q$ in varying the interaction strength from purely repulsive to weakly-attractive indicating that by tuning the R/L-CP interaction one can vary the domain size of the lamellae without changing the chain length. From the distribution of chain size in the film, see figure~\ref{fig: rg-pdf}, we conclude that the domain size variation is a consequence of the variation of size of L-CPs with changing interaction strength. 
On the other hand, chain conformation studies reveal that on average R-CPs are relatively more compact and observed threading of linear chains through rings which will have consequences in the dynamics (not addressed in this study). And as revealed from the data of shape parameters we find that the average instantaneous shapes of R-CPs and L-CPs are very different, see figure~\ref{fig: ellipsoid}, and relative to the chains in bulk both ring and linear chains are elongated and looks more flat near the walls, but only ring CPs assumes an oblate shape. 
Furthermore, from the distribution of chains in the film we found the segregation of R-CPs near the walls and we attribute this segregation due to the entropic force, i.e., due to the more compact structure of R-CPs in the bulk the depletion force sets in and hence drives it to the interface. \
In this this we have considered only 50:50 mixture and one has to consider other compositions. Also, it would be interesting to see a detail analysis on the internal structure of the copolymer rings, e.g. by calculating the average distance between monomer $i$ and $j$, nature of swelling at various degree of polymerization, and dynamics of relatively short copolymer rings, and we leave this points for future work.


\begin{acknowledgements}
Support through DST-INSPIRE Faculty Award (grant number: DST/INSPIRE/04/2015/001914) is gratefully acknowledged. 
\end{acknowledgements}


\end{document}